# Photo-physical and nonlinear-optical properties of a new polymer: hydroxylated pyridyl para-phenylene

**W. Ji [*], Hendry Izaac Elim, and Jun He,**

Department of Physics, National University of Singapore,

2 Science Drive 3, Singapore 117542, Singapore

**F. Fitrilawati, C. Baskar, S. Valiyaveettil*, and W. Knoll**

Temasek Professorship Program, Departments of Chemistry and of Materials

Science, National University of Singapore,

3 Science Drive 3, Singapore 117543, Singapore

**Abstract**

Photo-physical and nonlinear-optical properties of a new amphiphilic conjugated polymer: hydroxylated pyridyl para-phenylene (Py-PhPPP) both in $CH_2Cl_2$ solution and in thin films have been investigated. By using Z-scan technique with nanosecond laser pulses of wavelengths ranging from 430 to 600 nm, the large nonlinear absorption and refraction have been determined in terms of the effective third-order, nonlinear-optical susceptibilities. These Z-scans reveal that the nonlinear absorption alters from reverse saturable absorption to saturable absorption at wavelength of ~ 540 nm. Similarly, alteration from self-defocusing to self-focusing manifests itself at the same wavelength. The optical limiting performance of Py-PhPPP in solution is superior to toluene solution of [60]fullerene ($C_{60}$) at 532 nm. Both UV-visible absorption spectra and photoluminescence (PL) spectra show concentration dependence. The PL spectra also depend on excitation wavelengths. These evidences suggest that the aggregate formation should play an important role in the nonlinear-optical properties of the new polymer. We attribute the reverse saturable absorption in the region of blue and green wavelengths mainly to intra-chain, triplet-triplet absorption while the absorption bleaching at longer wavelengths is due to saturation in the absorption band induced by the aggregates.

---

[*] Corresponding author, email address: *phyjiwei@nus.edu.sg*

1.   **Introduction**

Recently there has been an enormous interest in organic conjugated polymers because they possess strong luminescence, a great potential for light emitting diodes or lasers [1]. These polymers are also known to exhibit large, ultra-fast third-order optical non-linearity at red and near-infrared wavelengths, promising for all-optical switching required by optical communications [2-13]. Among these polymers, a class of poly(para-phenylene)-type ladder-polymers (LPPPs) is one of the most attractive polymers. It is now widely accepted that, upon photo-excitation, LPPPs at its ground state, $S_0$ are excited to its lowest-lying singlet exciton, $S_1$. After rapidly relaxing to the bottom states of $S_1$, a part of the excited electrons radiatively recombine with $S_0$, while another part may transfer to the states of the lowest-lying triplet exciton, $T_1$, through inter-system crossing, as illustrated in Fig. 1. The nature of the $S_0 - S_1$ transition in LPPPs has been intensively investigated [4, 11-13]. However, excited state absorption (or photo-induced absorption), has not been fully understood yet in LPPPs, with excitation of laser pulses on the nanosecond scale, in which triplet-triplet transitions may make significant contribution. The triplet excited state absorption may result in reverse saturable absorption if the absorption cross section of the $T_1 - T_n$ transition is greater than that of the $S_0 - S_1$ transition. It can be exploited for protection of optical sensors or the human eye from intense laser radiation. The best-known reverse saturable absorbers are [60]fullerene ($C_{60}$) [14], and phthalocyanine complexes [15].

Here we report our research on the nonlinear absorption and nonlinear refraction in a new amphiphilic conjugated polymer: hydroxylated pyridyl para-phenylene (Py-PhPPP) with nanosecond laser pulses. The new polymer has a modified structure from that of LPPP or PPP, as displayed in Fig. 2. We shall demonstrate that this new polymer possesses strong reverse saturable absorption both in $CH_2Cl_2$ solution and in thin film. Our results show that the polymer is an excellent optical limiter in the blue and green spectral regions. In addition, we find that polymer aggregates play an important role at red and near-infrared wavelengths, and their absorption saturation leads large optical non-linearities.

## 2. Experimental

The new amphiphilic conjugated polymer, pyridine-incorporated polyhydroxy-(polyparaphenylenes) (Py-PhPPP), was synthesized by Suzuki poly-condensation under standard conditions as described earlier [16]. The polymer showed good solubility in common organic solvents such as chloroform, dichloromethane, tetrahydrofuran (THF), toluene, dimethylformamide (DMF), and formic acid. The PyPhPPP polymer has a molecular weight $M_w$ = 4241, and a glass transition temperature $T_g$ = 125 °C. Polymer solutions for nonlinear-optical measurements were prepared using $CH_2Cl_2$ as a solvent. Thin films of Py-PhPPP were prepared by spin-coating a solution of Py-PhPPP in toluene on fused silica substrates (Spectrosil 2000), followed by thermal annealing at 60°C for approximately 12 h in vacuum to remove the residual solvent. The film thickness was measured to be ~200 nm with a surface profiler (Alpha-Step 500).

The linear transmission spectra of the new polymer were obtained with a spectrophotometer (Shimadzu, UV-1601). The linear absorption coefficients, $\alpha_0$, were calculated from the transmission spectra of the Py-PhPPP ultra-thin film after correction of reflection losses at film–air and film–substrate interfaces. The dispersion of the linear refractive index, $n_0$, of the films was evaluated from the transmission and reflection spectra, measured at nearly perpendicular incidence, using the Kramers-Kronig formulation [17]. The photoluminescence (PL) spectra of the polymer were observed by using a luminescence spectrophotometer (Perkin-Elmer Instrument, LS 55) with excitation at 380-nm or 440-nm wavelength.

The nonlinear-optical properties of the polymer were investigated by both Z-scan and optical-limiting measurements with linearly polarized laser pulses of 5-ns or 7-ns duration from an optical parametric oscillator (Spectra-Physics, MOPO 710), or a Q-switched, frequency-doubled Nd:YAG laser (Spectra-Physics, DCR3), respectively. The spatial distribution of the pulses was nearly Gaussian after passing through a spatial filter. The pulse was divided by a beam splitter into two parts. The reflected part was taken as the reference representing the incident pulse energy and the transmitted beam was focused through the sample. Both the incident and the transmitted pulse energies were measured simultaneously by two pyro-electric detectors (Laser Precision, RjP-735). The minimum beam waist of the focused laser beam was ~28 μm, determined by the standard Z-scan method [18]. To conduct the Z-scans, the sample was moved along the laser light propagation direction while both the incident and the transmitted pulse energies were recorded. The optical-limiting measurements were carried out when the sample was fixed at the focal point.

## 3. Results and discussion

### A. *UV-Visible absorption and its concentration dependence*

Figure 3 displays the absorption spectra of Py-PhPPP in both thin film and solutions. These spectra show well-structured bands indicative of a high degree of conjugation and high intra-chain order. For the film, the absorption peak is located at 409 nm with a maximum absorption coefficient $\alpha_0$ = 6 x $10^4$ $cm^{-1}$. Relative to LPPP [4], it has a blue shift of ~ 50 nm. For comparison, Fig. 3(a) also shows the absorption spectrum of the PhPPP film (dotted line) coated on quartz substrate. The overall shape of Py-PhPPP (incorporated with pyridine) is similar to that of PhPPP (without pyridine), but with a red shift [16]. The observed shift indicates that Py-PhPPP contains on-average, longer conjugated segments than those chains in PhPPP, and hence, a larger density of de-localized $\pi$-electrons gives rise to a red shift.

It is worthwhile noticing that the observed red shift in the absorption peak positions from the dilute solution to the film, indicative of the strong dependence on the polymer concentration, as illustrated by the inset of Fig. 3(a). For a dilute solution (0.1 mg/ml), the polymer chains are presumably isolated, the absorption peak at 370 nm is dominated by the intra-chain singlet exciton. As the Py-PhPPP concentration in the solution is raised, the absorption shifts to red, implying that a new absorption band appears to the red side of the normal absorption band of the intra-chain singlet exciton. Such an absorption band has been also observed in MEH-PPV polymer [19], polypyridines (PPy) and pyridine incorporated polymers, PPyV and PPyVPV [20]; and attributed to the formation of aggregates. The aggregation provides more chain segments in contact over which the electron wave-function can be de-localized, leading a redder ground state absorption, (and it is referred to as the inter-chain exciton). We expect that such inter-molecular interactions reach a maximum in the film. The formation of aggregates leads to an absorption tail spanning the entire visible region observed in the film, partially due to the inter-chain absorption band and partially caused by Rayleigh scattering. The refractive index of the polymer film decreases from 1.97 to 1.63 as the wavelength increases from 410 nm to 800 nm, as shown in Fig. 3(b).

## B. *Photoluminescence, its concentration and excitation dependence*

It is anticipated that the photo-physics in Py-PhPPP in the range from 300 nm to 2 μm is determined by a series of alternating odd- ($B_u$) and even- ($A_g$) parity excited states [21-22], corresponding to one-photon and two-photon allowed transitions, respectively. Optical excitation into either of these states is followed by subpicosecond, non-radiative relaxation to the lowest excited state [23]. This relaxation is due to either vibrational cooling within vibronic side-bands of the same electronic state or phonon-assisted transitions between two different electronic states. In molecular spectroscopy, the latter process is referred to as internal conversion. Internal conversion is usually the fastest relaxation channel, providing efficient non-radiative transfer from a higher excited state to the lowest excited state of the same spin multiplicity. As a result, the vast majority of molecular systems follow the Vavilov-Kasha rule, stating that fluorescence associated with photon absorption typically occurs from the lowest excited electronic state and its quantum yield is independent of the excitation wavelength [24]. The Stokes shift between the absorption ($\lambda_{max}$ = 370 ~ 380 nm) and PL spectra ($\lambda_{max}$ = 490 ~ 510 nm), as shown in Fig. 4 for solutions of Py-PhPPP, is consistent with this rule.

The concentration dependence of the PL spectra also gives direct evidence for the formation of aggregates in the solutions of Py-PhPPP. Fig. 4(a) displays a red-shift in the PL peak position from 490 nm to 510 nm when the Py-PhPPP concentration is increased from 0.5 to 1 mg/ml. As expected, the higher concentration of Py-PhPPP, the greater the degree of aggregates. Hence the emission band is enhanced in the red wavelengths due to the addition of the inter-chain to the intra-chain exciton recombination. The inter-chain emission band has been observed previously in LPPP [25-26] and MEH-PPV polymers as well [19].

Fig. 4(b) shows the excitation dependence of the PL spectra obtained when the concentration is kept the same but the excitation wavelength is changed from 380 nm to 440 nm. Note that the 380-nm excitation is nearly resonant with the intra-chain singlet exciton, while the 440-nm excitation is close to the red edge of the main absorption band. With the 380-nm excitation, therefore, the PL spectrum is expected to be dominated by the recombination of the intra-chain exciton. However, for excitation at 440 nm, the inter-chain exciton recombination becomes more pronounced, thus the PL spectrum is red shifted. Furthermore, the PL quantum yield is decreased when the excitation wavelength is changed from 380 nm to 440 nm. Such a quenching normally happens at high excitation densities for the intra-chain exciton. Here we interpret it in terms of the formation of different species as the inter-chain exciton recombination overplays its intra-chain counterpart.

### C.     *Z-scan measurements*

The nonlinear-absorptive and nonlinear-refractive properties of Py-PhPPP in $CH_2Cl_2$ are illustrated in Fig. 5, measured with the open- and closed-aperture Z-scans at wavelengths ranging from 430 nm to 600 nm. Similar behavior is also observed in the PhPPP and Py-PhPPP films. Figure 6 not only illustrates an example but also shows an expected improvement to the photo-induced absorption due to the incorporation of pyridine to PhPPP backbone. The open-aperture Z-scans in Fig. 5(a) clearly demonstrate the reverse saturable absorption at shorter wavelengths ($\lambda < 540$ nm), while at longer wavelength ($\lambda > 540$ nm), a weak photo-induced bleaching manifests itself in the Z-scans. Similarly, the change in the sign of $n_2$ is shown in Fig. 5(b). We observe a negative $n_2$ (self-defocusing) at $\lambda < 540$ nm. At approximately 540 nm, $n_2$ is zero and becomes positive (self-focusing) at longer wavelengths.

We assume that both photo-induced absorption and refraction can be described by $\Delta\alpha = \beta\, I$, and $\Delta n = n_2\, I$, where $\beta$ and $n_2$ are the nonlinear absorption coefficient and nonlinear refractive index, respectively, and $I$ is the light intensity. By fitting to the Z-scan theory [18], we extract the largest values of $\beta$ and $n_2$ near the absorption peak at 430 nm with $\beta = 2500$ cm/GW and $n_2 = -7.5 \times 10^{-3}$ $cm^2$/GW. At 600 nm, the measured $\beta$ and $n_2$ values are $-33$ cm/GW and $0.25 \times 10^{-3}$ $cm^2$/GW, respectively. Figure 7 shows the irradiance independence of the observed nonlinear coefficients, confirming that our assumption of $\Delta\alpha = \beta\, I$ and $\Delta n = n_2\, I$ is justified. Table 1 lists all the measured nonlinear parameters with two figures of merits commonly used to assess the polymer for all-optical switching.

For most organic polymers, the nonlinear absorption are normally expected to originate from the photo-physical processes such as two-photon transitions of singlet excitons, singlet excited state absorption, and triplet excited state absorption, occurred within the intra-chains. However, as indicated by our absorption and PL spectra, we expect that the photo-dynamics due to the aggregates in the Py-PhPPP should be important too. The observed nonlinear effects are the result of interplay between these processes. It has been reported that excited electrons can be transferred from singlet to triplet states on picosecond time scale for pyridine-based polymers [20]. With the nanosecond pulses employed in our experiments, we speculate that the photo-induced absorption at shorter wavelengths (<540 nm) is dominated by the intra-chain, triplet $T_1$-$T_n$ absorption, although both singlet excited state absorption and two-photon absorption may contribute. It should be pointed out that the exact dynamic processes can be revealed by using femtosecond time-resolved, pump-probe measurements [19,20]. At longer wavelengths, however, the saturation of the inter-chain singlet exciton becomes important.

In general, reverse saturable absorption may be explained quantitatively in terms of the five-level model [15]. For nanosecond laser pulses and strong triplet-triplet absorption, it can be approximately described by a sequential two-photon absorption (STPA) process, defined as $\alpha = \alpha_0 + \beta_{eff} I$; where $\beta_{eff}$ is the effective third-order nonlinear absorption coefficient, describing the intersystem crossing time, ground-state ($S_0$-$S_1$) and triplet ($T_1$-$T_n$) absorption cross sections. Our Z-scan results confirm strong triplet excited state absorption in Py-PhPPP in the blue and green wavelength regions. For the absorption bleaching due to saturation of the inter-chain singlet exciton, a simple two-level model predicts $\text{Im}[\chi^{(3)}] \sim -1/[(\omega-\omega_0)^2 + \Gamma^2]$, explaining the negative sign of the measured effective $\beta$.

Similarly, the observed nonlinear refraction is believed to be the result of saturation on the intra- or inter-chain effects. For short wavelengths, the self-defocusing results from the saturation of the intra-chain, singlet exciton, predictable by the two-level model: $Re[\chi^{(3)}] \sim (\omega-\omega_0)/[(\omega-\omega_0)^2 + \Gamma^2]$ when $\omega < \omega_0$, (or $\lambda > \lambda_{max} = \sim 400$ nm). At longer wavelengths, however, the saturation of inter-chain singlet exciton becomes dominant. The two-level model gives a positive sign for the nonlinear refraction when $\omega > \omega_0$, (or $\lambda < \lambda_{max} = \sim 600$ nm expected for the inter-chain exciton,) in agreement with our measurements.

### D.  *Optical limiting*

The observed strong excited state absorption can be exploited for optical-limiting applications. Figure 8 shows that the energy-dependent transmission of 1.0 mg/ml Py-PhPPP in solution is a constant until the input energy of $\sim 0.7$ J/cm$^2$ with 532-nm, 7-ns laser pulses. However, when the input energy increases beyond $\sim 0.7$ J/cm$^2$, the measured transmittance deviates from the linearity and decreases dramatically at $\sim 2$ J/cm$^2$, indicating the occurrence of optical limiting. For comparison, a $C_{60}$-tolune solution with the same linear transmittance ($\sim 90\%$) has been tested in the same experimental set-up, showing poorer limiting behavior, in particular, in the high-energy regime. Note that the poor limiting behavior observed in our experiment for $C_{60}$ is due to the high linear transmittance used, whereas most reported results for $C_{60}$ to have better limiting performance are achieved with the linear transmittance of 70% or less [14]. We should emphasize the following points. (1) Self-defocusing of Py-PhPPP has not been fully exploited in our limiting measurements since it has been carried out without placing aperture in front of the transmission detector. (2) Our Z-scans indicate that we should observe much stronger limiting behavior when the laser wavelength is tuned to shorter wavelengths. And (3) to check photostability of all the samples, we have measured and compared the absorption spectra before and after laser irradiation. The obtained results indicate that there is no difference in the spectra for all the samples, showing that all the samples have a good photostability.

4. **Conclusion**

The linear- and nonlinear-optical properties of a new amphiphilic conjugated polymer: hydroxylated pyridyl para-phenylene (Py-PhPPP) both in $CH_2Cl_2$ solution and coated on quartz substrate have been investigated. Its UV-visible absorption spectra show a dominant absorption peak at ~400 nm and it is concentration dependant. By using Z-scan technique with nanosecond laser pulses of wavelengths ranging from 430 to 600 nm, the observed large nonlinear absorption and refraction have been determined in terms of the effective third-order nonlinear-optical susceptibilities. The Z-scans reveal that the nonlinear absorption alters from reverse saturable absorption to saturable absorption at wavelength of ~540 nm. Similarly, alteration from self-defocusing to self-focusing manifests itself at the same wavelength. The optical limiting perforamnce of Py-PhPPP in solution is superior to the toluene solution of [60]fullerene ($C_{60}$) at 532 nm. The photoluminescence (PL) spectra of the polymer show concentration dependence. And the PL spectra also depend on excitation wavelength. These evidences suggest that the aggregates should play an important role in the new polymer. We attribute the reverse saturable absorption in the region of blue and green wavelengths mainly to intra-chain, triplet-triplet absorption while the absorption bleaching at longer wavelengths is due to saturation in the absorption band induced by the aggregates.

**Acknowledgements**

We thank the National University of Singapore and Temasek Professorship Program for financial supports of this work.

**Table 1** Measured linear and nonlinear absorption coefficient ($\alpha_0$ and $\beta$), nonlinear refractive index ($n_2$), third-order susceptibility $|\chi^{(3)}|$, and figures of merit for Py-PhPPP in $CH_2Cl_2$ solution with concentration of 1.0 mg/ml. The irradiance, $I$, used for calculation of $W$ is 0.44 GW/cm$^2$.

| $\lambda$ (nm) | $\alpha_0$ (cm$^{-1}$) | $\beta$ (x 10$^2$ cm/GW) | $n_2$ (x 10$^{-3}$ cm/GW) | $|\chi^{(3)}|$ (x 10$^{-10}$ esu) | $T = |\beta\lambda/n_2|$ | $W = |n_2 I/\alpha_0\lambda|$ |
|---|---|---|---|---|---|---|
| 430 | 31.4 | 25 | -7.5 | 13.8 | 14.33 | 2.4 |
| 450 | 9.7 | 16 | -6.2 | 9.0 | 11.61 | 6.3 |
| 470 | 4.6 | 1.2 | -1.1 | 5.2 | 5.13 | 2.2 |
| 500 | 1.8 | 0.17 | -0.55 | 0.41 | 1.54 | 2.7 |
| 560 | 0.6 | -0.24 | 0.15 | 0.15 | 8.96 | 2.0 |
| 600 | 0.4 | -0.33 | 0.25 | 0.24 | 7.92 | 5.2 |

*Figure captions*

**Fig. 1** Schematic representation of photo-dynamics in conjugated polymers. $S_0$ is the ground (singlet) state; $S_1$ is the first excited singlet state; $T_1$ is the first excited triplet state; and $T_i$ and $T_n$ are higher-lying excited triplet states. The solid lines represent fluorescence (a), intersystem crossing (b), triplet-triplet absorption (c), and phosphorescence (d). The dashed lines denote non-radiative processes.

**Fig. 2** Molecular structure of Py-PhPPP.

**Fig. 3 (a)** UV-visible absorption spectra of PhPPP (dotted line) and Py-PhPPP (solid line) films with the film thickness of ~200 nm. The inset shows the absorption spectra of 1.0 mg/ml Py-PhPPP (solid line), 0.5 mg/ml Py-PhPPP (dashed line), and 0.1 mg/ml Py-PhPPP (dotted line) in $CH_2Cl_2$ solution. Note that these spectra are normalized to their peaks. **(b)** The calculated absorption coefficient (dotted line) and refractive index (solid line) as a function of the wavelength.

**Fig. 4 (a)** PL spectra of 0.5 and 1.0 mg/ml Py-PhPPP in $CH_2Cl_2$ solution excited at a wavelength of 440 nm. **(b)** PL spectra of 1.0 mg/ml Py-PhPPP in $CH_2Cl_2$ solution excited at wavelengths of 380 nm and 440 nm.

**Fig. 5 (a)** Open-aperture and **(b)** closed-aperture Z-scan measurements performed on 1.0 mg/ml, 1-mm-thick Py-PhPPP in $CH_2Cl_2$ solution at different wavelengths. All the solid lines are the theoretical fits by using the Z-scan theory [18]. All the Z-scans are conducted with a beam waist of 28 μm, a pulse repetition rate of 10 Hz and a peak irradiance of 44 $MW/cm^2$. Some of the Z-scans are vertically shifted for clear presentation.

**Fig. 6** Open-aperture Z-scan measurements performed on the 200-nm-thick PhPPP and Py-PhPPP films at 470 nm. The experimental conditions are the same as those described in Fig. 5. The solid lines are fittings using the Z-scan theory [18].

**Fig. 7** Irradiance independence of the nonlinear absorption coefficient (filled circles) and the nonlinear refractive index (open circles) in the 1.0 mg/ml, Py-PhPPP/$CH_2Cl_2$ solution observed at 500 nm.

**Fig. 8** Transmittance of the 1 mg/ml, 1-mm-thick $CH_2Cl_2$ solution of Py-PhPPP (open circles) and the 1-mm-thick $C_{60}$/toluene solution (filled circles) measured as a function of the input fluence at 532 nm. The linear transmittance of the two solutions is ~ 90%.

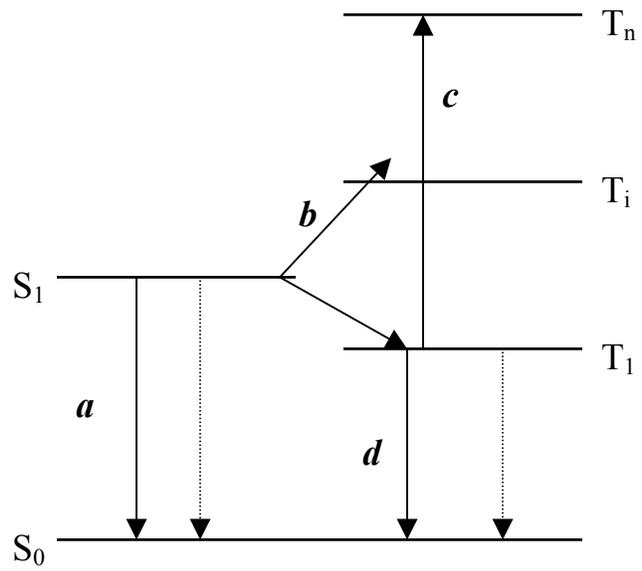

**Figure 1.**

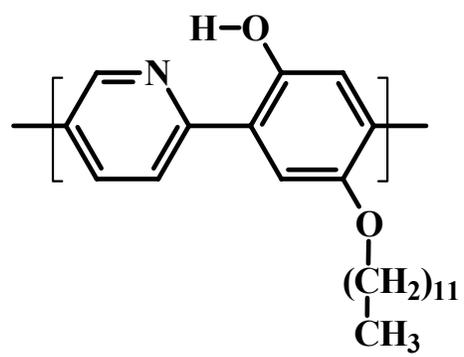

**Figure 2.**

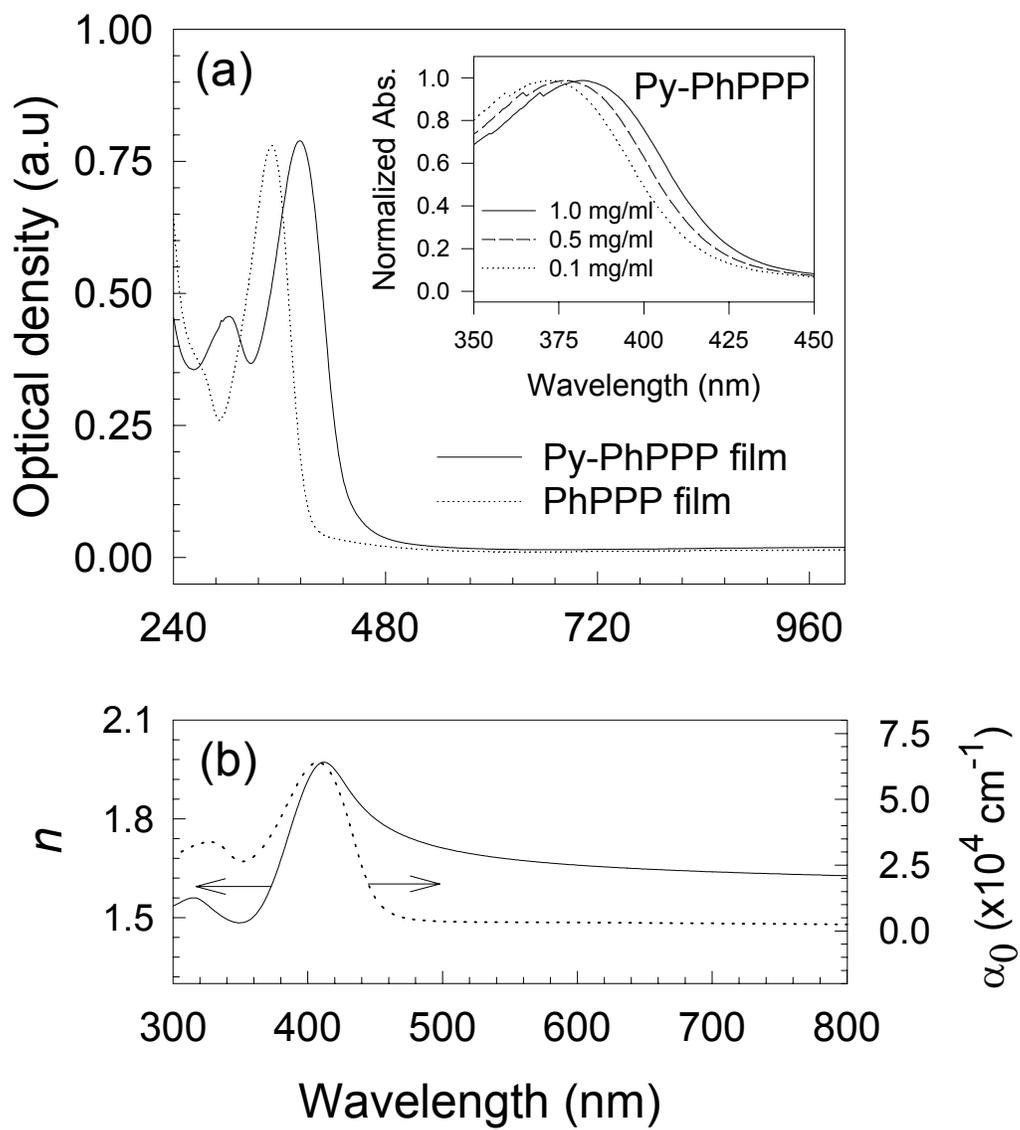

**Figure 3.**

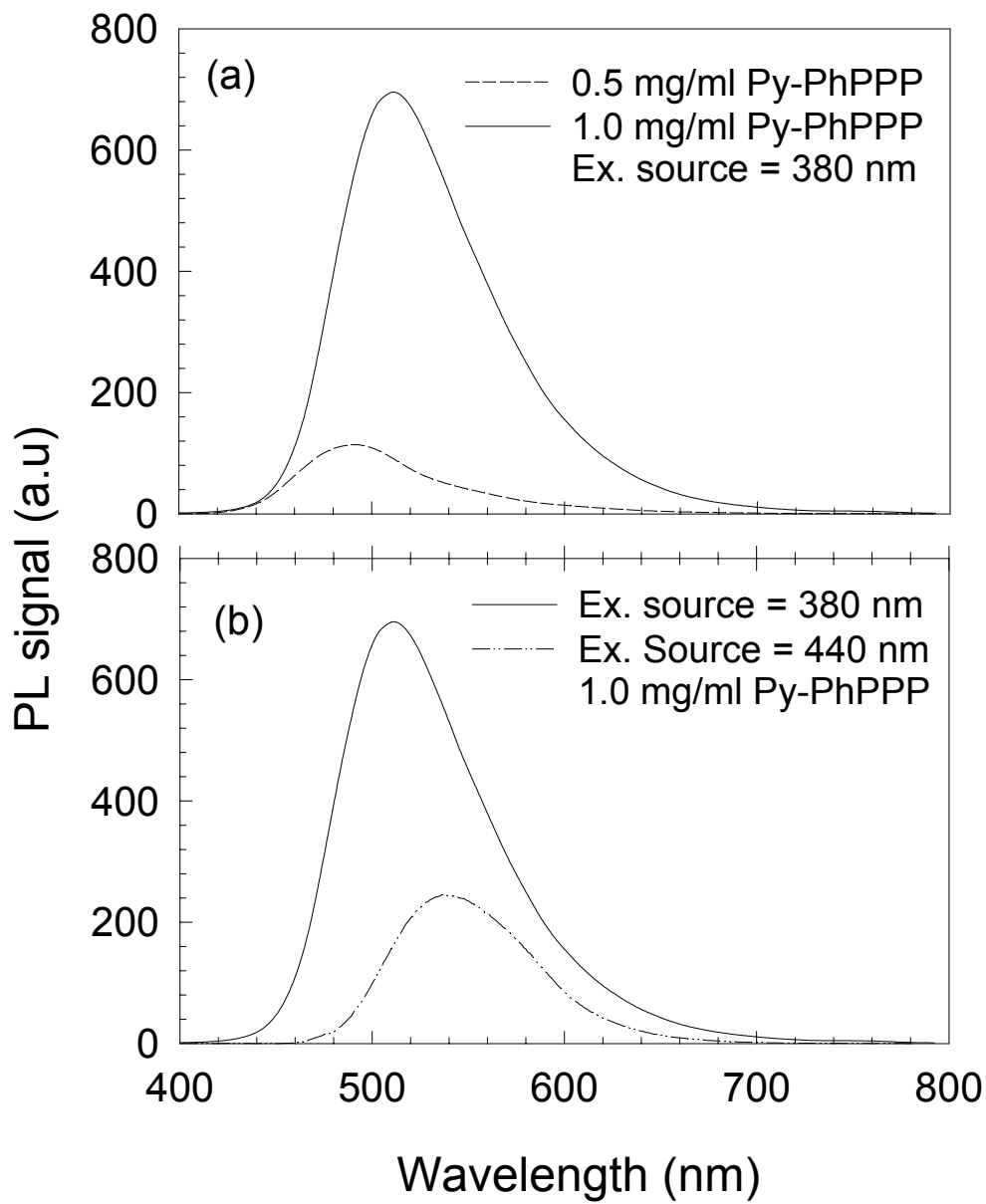

**Figure 4**

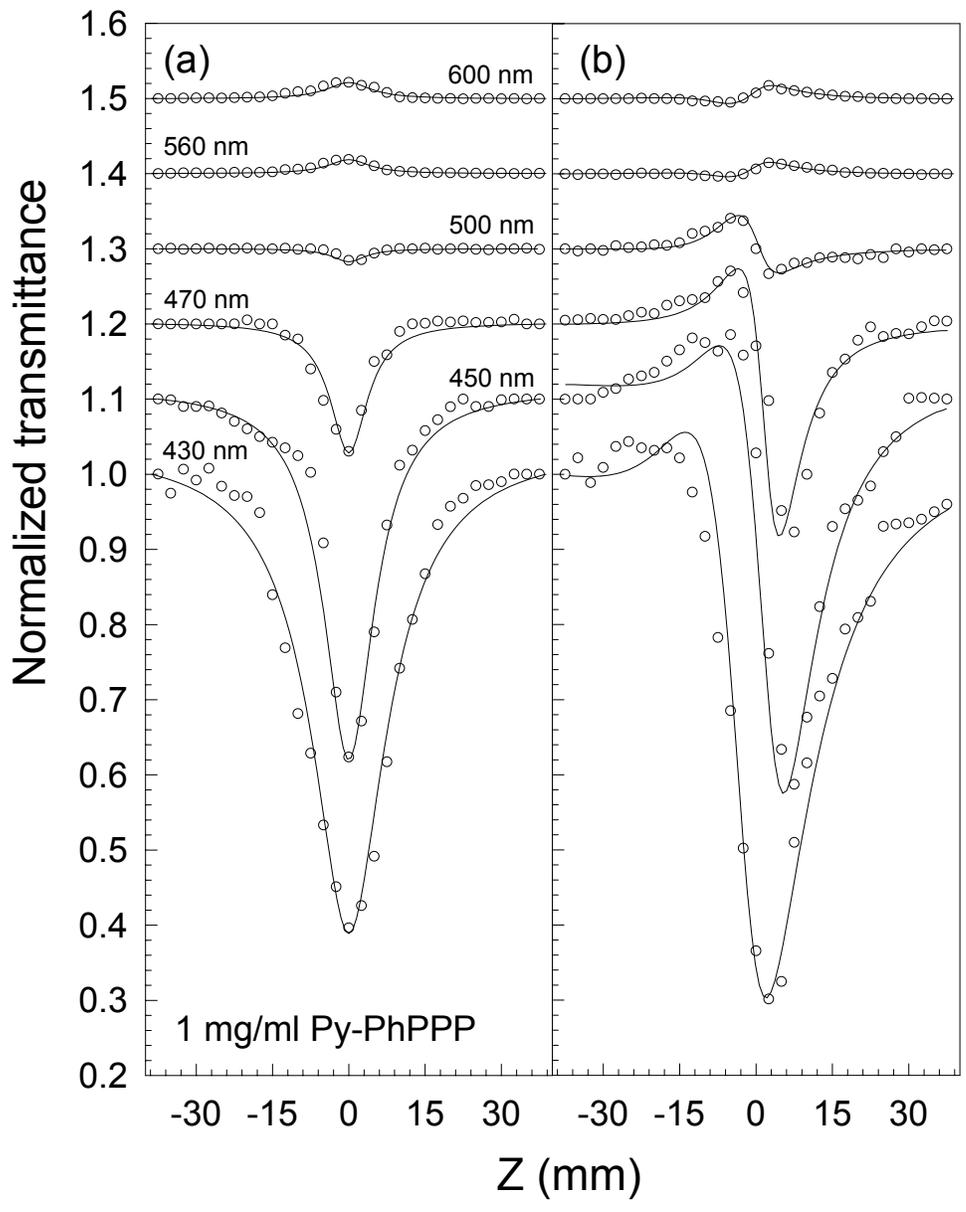

**Figure 5.**

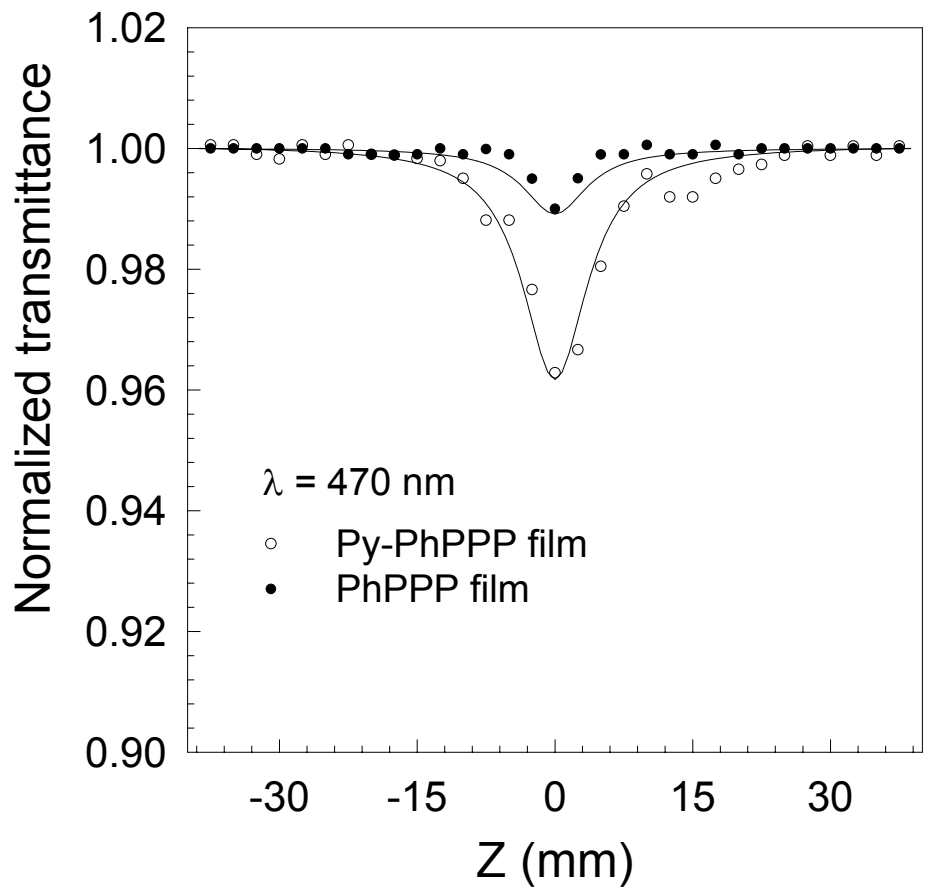

**Figure 6.**

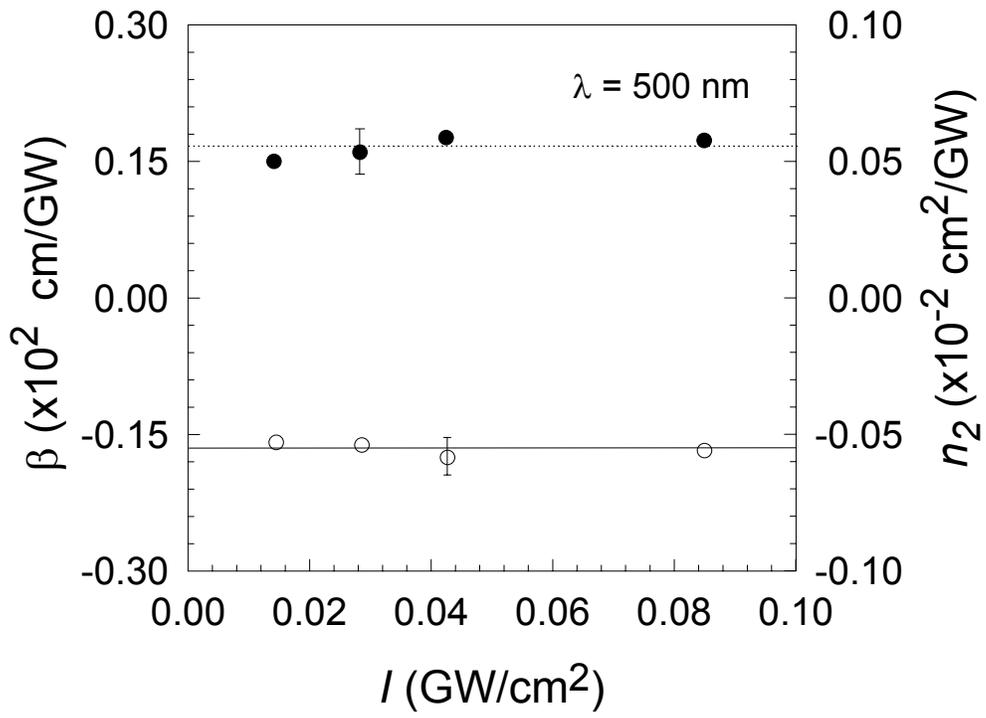

**Figure 7.**

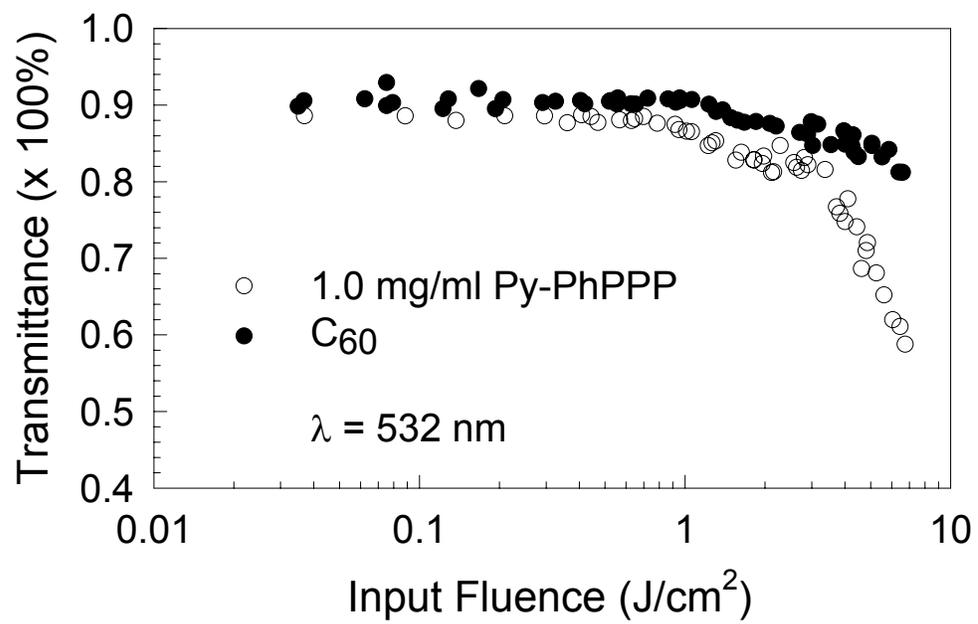

**Figure 8**